\documentclass[aps,pre,showkeys,preprint,superscriptaddress]{revtex4-1}
\usepackage{natbib}
\usepackage{epsf}
\usepackage{amsmath,amssymb,graphicx}

\begin{document}

\title{Rotating Shallow Water Dynamics:\\
Extra Invariant and the Formation of Zonal Jets.}

\author{Alexander M. Balk}
\affiliation{Department of Mathematics, University of Utah,
155 South 1400 East, Salt Lake City, Utah 84112, USA}
\author{Francois van Heerden}
\affiliation{Nuclear Energy Corporation of South Africa,
Reactor and Radiation Theory, Building P-1900, POBox 582, Pretoria
0001, South Africa}
\author{Peter B. Weichman}
\affiliation{BAE Systems,
Advanced Information Technologies,
6 New England Executive Park,
Burlington, Massachusetts 01803, USA}

\renewcommand{\theequation}{\thesection.\arabic{equation}}

\date{\today}

\begin{abstract}

We show that rotating shallow water dynamics possesses an approximate
(adiabatic-type) positive quadratic invariant, which exists not only at
mid-latitudes (where its analogue in the quasigeostrophic equation has
been previously investigated), but near the equator as well (where the
quasigeostrophic equation is inapplicable). Deriving the extra
invariant, we find ``small denominators'' of two kinds: (1) due to the
triad resonances (as in the case of the quasigeostrophic equation) and
(2) due to the equatorial limit, when the Rossby radius of deformation
becomes infinite. We show that the ``small denominators'' of
\emph{both} kinds can be canceled. The presence of the extra invariant
can lead to the generation of zonal jets. We find that this tendency
should be especially pronounced near the equator. Similar invariant
occurs in magnetically confined fusion plasmas and can lead to the
emergence of zonal flows.

\end{abstract}

\keywords{Rossby waves; Drift waves;
Triad resonance; Shallow water;
Conservation; Adiabatic invariants;
Zonal jets; Zonal flows in plasmas}


\maketitle

\section{Introduction}
\setcounter{equation}{0}
\label{Sect:Intro}

To introduce the topic of present paper, let us start with two
different physical situations, which are known.

The first is the inverse cascade of energy in two-dimensional
hydrodynamics \cite{BatchHomTurb,Fort}. The inverse cascade is related
to the presence of an additional (compared to the 3D hydrodynamics)
positive quadratic invariant --- {\it enstrophy}.

The second situation \cite{ZakhSurf} is the appearance of longer waves
in sea-wave turbulence: The length of typical waves on the sea surface
is often much bigger than those generated directly by the wind and
increases with time (a process known as \emph{wave aging}). The sea
waves generated directly by wind produce---via nonlinear
interaction---longer waves, the latter produce even longer waves, and
so on. This process is a manifestation of the inverse cascade
\cite{Zakh68}. As in the first situation, the inverse cascade is
related to the presence of another (in addition to the energy) positive
quadratic invariant, in this case the {\it wave action}. The later
invariant holds because the gravity wave dispersion forbids 3-wave
interactions, and so the main resonance interaction involves 4 waves
and conserves the total wave action. The conservation of the wave
action is similar to the conservation of the total number of molecules
in rarefied gas (when the main interactions are binary collisions).
However, unlike the number of molecules, the wave action is only an
approximate invariant, whose conservation fails in higher order
interactions (e.g., 5-wave interactions are possible, and they fail to
conserve the wave action).

The present paper considers wave dynamics in rapidly rotating
geophysical fluids. An integral part of this dynamics is the emergence
of zonal jets \cite[][]{Rhi}, see also the collection of papers
\cite[][]{Jets}; the stripes on Jupiter make a famous example
\cite{NASA}. Zonal jets  are, as well, observed in the dynamics of
magnetized plasmas (which is mathematically similar to geophysical
fluid dynamics); they appear to act as transport barriers in tokamaks
\cite{Diamond}. Plasma regimes with zonal jets have become an integral
part of modern controlled nuclear fusion installations, in particular,
ITER.

It is interesting to see if the emergence of zonal jets can be related
to the existence of an additional invariant (similar to the two
examples given above). This is the main motivation of the present
paper.

It is believed that zonal jets emerge as a result of Rossby wave
interactions, see \cite{Newell,Rhines}. The nonlinear dynamics of
Rossby waves in the beta plane is often modeled by the quasigeostrophic
equation (see e.g. \cite[][]{Slmn})
\begin{eqnarray}
(\Delta \psi - \alpha^2 \psi)_t
+ \beta \psi_x + \psi_y \Delta \psi_x - \psi_x \Delta \psi_y = 0
\label{CHM}
\end{eqnarray}
for the stream function $\psi(x,y,t)$ of the horizontal fluid velocity
$(u,v) = (\psi_y,-\psi_x)$. Here, $\alpha$ is the inverse Rossby radius
of deformation, and $\beta$ is the beta parameter characterizing the
variation of the Coriolis force. The subscripts $x,y,t$ denote partial
derivatives, and $\Delta$ is the two-dimensional Laplacian.

It has been shown that the quasigeostrophic equation (\ref{CHM}) indeed
possesses an approximate (adiabatic-type) quadratic invariant, which
requires the inverse cascade to transfer energy not just to large
scales but specifically towards zonal flow
\cite[][]{BNZ,B1991,B2005,Nazarenko,Nazarenk}.

However, in several physical situations (including Jupiter) zonal jets
are well pronounced near the equator, while equation (\ref{CHM}) is not
applicable there. It is a major challenge to see if the approach to
zonal jets based on the extra invariant works near the equator as well.

To account for the equatorial region, we consider the rotating shallow
water dynamics in the beta-plane (equatorial or mid-latitudinal)
\begin{subequations} \label{RSW}
\begin{eqnarray}
u_t \,&+&\, u\, u_x\, + \,v\, u_y\, - \,f(y)\, v\, =\, -g\, H_x\,,\label{RSWu}\\
v_t \,&+&\, u\, v_x\, + \,v\, v_y\, + \,f(y)\, u\, =\, -g\, H_y\,,\label{RSWv}\\
H_t \,&+&\, (H\,u)_x\, + \,(H\,v)_y\, =\, 0\,,                    \label{RSWH}
\end{eqnarray}
\end{subequations}
e.g. \cite{Va,McW}. This system of equations describes the evolution of
the horizontal fluid velocity $(u,v)$ and the fluid height $H$ (flat
bottom is assumed). The function $f(y)$ is the Coriolis parameter, and
$g$ is the acceleration due to gravity.

Considering perturbation expansions (see below) for the system
(\ref{RSW}), we find ``small denominators'' not only related to the
resonance triads [like in the case of (\ref{CHM})], but also related to
the equatorial limit ($f \rightarrow 0$). We show that ``small
denominators'' of both kinds can be canceled. The possibility of such
cancelation is a remarkable property of the rotating shallow water
system. We are unaware of any other system with similar attributes,
even remotely.

Once we pass from a single equation (\ref{CHM}) with constant
coefficients to a system (\ref{RSW}) of three equations with
$y$-dependent coefficients, we also face two other problems:
\begin{enumerate}
\item There could be resonant interactions and energy transfer
    between the Rossby waves and the inertia-gravity waves.

\item The translational symmetry is broken.
\end{enumerate}
The first problem is resolved due to a general fact observed in a
variety of rotating fluid systems: The coupling constant in the triad
interaction between a slow mode and two fast modes vanishes in the
equation for the slow mode \cite{Chen2EyinkHolm}; in our case, the slow
is the Rossby mode, and the fast is the inertia-gravity mode. The
second problem makes perturbation expansions significantly harder; in
particular, the Rossby mode needs to be refined (Section
\ref{Sect:mode}).

We show that in the limit of weak nonlinearity, the system (\ref{RSW})
possesses an additional approximate (adiabatic-type) invariant, which
is described in Sec.\ \ref{Sect:Extra}. Before its formal derivation in
Sec.\ \ref{Sect:Adiabatic}, we demonstrate how the presence of this
invariant makes the inverse cascade \emph{anisotropic} and steers
energy toward zonal flow (Sec.\ \ref{Sect:Emerge}). Even more specific
features, observed in some experiments, are in agreement with the
proposed picture:

\begin{itemize}
\item Near the equator, the emergence of zonal jets is more
    pronounced than it is at mid-latitudes (Sec.\ \ref{Sect:Why}).

\item In the opposite limit (when typical length of waves excited
    by forcing is much greater than the Rossby radius of
    deformation), the extra invariant just says that the energy
    should transfer into the sector of wave vectors ${\bf k}$ with
    polar angles $>60^\circ$ (Sec.\ \ref{Sect:LongW}).
\end{itemize}

During the last half century, several ideas were proposed to explain
the emergence of zonal jets through the dynamics of weakly nonlinear
Rossby waves, e.g., random wave closures \cite{Newell}, wave kinetic
equation \cite[][]{Rez}, modulational instability
\cite[][]{DiRoHiMaFlSm}, and almost resonant interactions
\cite[][]{LeeSmith}. Since these approaches consider the weakly
nonlinear regime, we believe, they should be intimately related to the
presence of the extra invariant. For the reasons discussed above, it is
crucial to see that this invariant is present not only in the
quasigeostrophic equation, but in the shallow water system as well.

To explain often powerful equatorial zonal jets, a deep approach was
developed over the past 40 years; it derives the formation of zonal
jets from the instability of the equatorial mixed Rossby-gravity waves
(see \cite{Shepherd,Fruman} and references therein). Since this
approach is also based on small nonlinearity, the presence of the extra
invariant remains relevant. We emphasize that, irrespective of the
detailed initial instability mechanism, it is the weak turbulent
inverse cascade that controls the geometry of any emergent large-scale
feature. Our theory then provides an underlying \emph{dynamical
mechanism} specific to the emergence of zonal jets, as observed in
numerical simulations. We do mention, however, the following issue that
merits future investigation. Very near the equator (within a couple of
degrees), there is evidence that dynamical terms neglected in the
``traditional approximation''  (in which the horizontal component of
Earth's angular velocity is ignored, producing the standard beta
plane approximation; see \cite{LeMy}) may play a significant role
\cite{Fruman}. Our theory ignores such terms, so it is presently
unknown if the extra invariant still exists under these conditions. The
present work, at minimum, extends the domain of validity
of the extra invariant from mid-latitudes to immediate vicinity of the
equator, where the quasigeostrophic equation is already inapplicable,
but the ``traditional approximation'' still remains valid.

We should also mention several mechanisms that connect zonal jets with
strong nonlinearity, when nonlinear terms (in dynamic equations) are
similar in magnitude to linear terms (see, e.g.,
\cite{Buhler,SkrnDGalp,Berloff}). Note that situations where linear
terms are \emph{negligible} compared to the nonlinear terms are not
related to the formation of zonal jets because only the linear terms
are anisotropic (the phenomenon of spontaneous emergence of anisotropy
is not relevant here because the emerging jets are always observed to
be parallel to the equator). As long as the linear terms are
significant, even if not dominant, the extra invariant should still
play an important role.

\section{Extra invariant}
\label{Sect:Extra}
\setcounter{equation}{0}

It is well known that (\ref{RSW}) conserves the following quantities:
\begin{itemize}

\item[$\diamond$] The energy, defined explicitly below by equation
    (\ref{E}).

\item[$\diamond$] The infinite set of potential vorticity
    integrals
\begin{align}
\label{PoVoSer}
	\int F\left[\frac{v_x-u_y+f(y)}{H}\right] \, H\, dxdy,
\end{align}
where $F$ is an arbitrary function of a single variable. These
conservation laws are related to the advective conservation of
potential vorticity \cite{Va}.

In particular, for $F \equiv 1$ one
obtains the total mass  $\int H\, dxdy$; its conservation implies
the existence of \emph{time-independent} space-averaged fluid
height $\bar H$, such that
\begin{align}
	\int (H-\bar H)\, dxdy = 0.
\end{align}

\item[$\diamond$]  The $x$-momentum
\begin{align}
	\int u\, H\, dxdy,
\end{align}
which is related to translational symmetry in zonal direction.

\end{itemize}
We will see that the dynamics (\ref{RSW}) \emph{adiabatically}
conserves three more quantities.

Before we describe them, let us eliminate some dimensional parameters
by rewriting (\ref{RSW}) in terms of the fractional relative height
\begin{eqnarray}\label{FrRelHeight}
h(x,y,t)=\frac{H-\bar H}{\bar{H}};
\end{eqnarray}
and rescaling
\begin{eqnarray}
ct \to t,\quad (u/c,v/c) \to (u,v),\quad f(y)/c \to f(y),
\quad\quad\mbox{where}\quad c = \sqrt{g \bar H}.
\end{eqnarray}
Then the shallow water dynamics (\ref{RSW}) takes the form
\begin{subequations}\label{uvh}
\begin{eqnarray}
u_t + u u_x + v u_y - f(y) v &=& - h_x,
\label{u} \\
v_t + u v_x + v v_y + f(y) u &=& - h_y,
\label{v} \\
h_t + u_x + v_y + (h u)_x + (h v)_y &=& 0,
\label{h}
\end{eqnarray}
\end{subequations}
where $u,v,h$ are dimensionless, while $x,y,t$ and $1/f$ have dimension
of length.

Consider the {\it linearized perturbational potential vorticity} (see \cite{Gill})
\begin{eqnarray}\label{PoVoL}
{\mathcal Q}=v_x-u_y-f(y)\,h\,,
\end{eqnarray}
which, according to (\ref{uvh}), obeys the equation of motion
\begin{equation}\label{Qt}
{\mathcal Q}_t\;+\;(u\,{\mathcal Q})_x\;
+\;(v\,{\mathcal Q})_y\;=\;-\beta\,v\,(1+h),\quad\mbox{ where}\quad \beta=f'(y).
\end{equation}
Since our goal is to describe the energy transfer in Fourier space, we
consider the Fourier transform ${\mathcal Q}_{\bf k}$ of the field
${\mathcal Q}$:
\begin{eqnarray}\label{Fourier}
{\mathcal Q}(x,y,t)=\int {\mathcal Q}_{\bf k}(t)\,
e^{i(p x + q y)}\, d p d q
\quad [{\bf k}=(p,q),\; k^2=p^2+q^2].
\end{eqnarray}

We will show that the shallow water dynamics (\ref{uvh}) adiabatically
conserves three quantities of the form
\begin{eqnarray}\label{TheIntegral}
I = \frac{1}{2}\int X_{\bf k} \;  {\mathcal Q}_{\bf k}\,
{\mathcal Q}_{-\bf k}\; dp\,dq.
\end{eqnarray}
The first is the energy of the Rossby waves (the inertia-gravity
component is excluded); it has
\begin{subequations}\label{kernels}
\begin{eqnarray}\label{energy}
X^{\mathrm{energy}}_{\bf k}=\frac{1}{f^2+k^2};
\end{eqnarray}
The second is the enstrophy of the Rossby component; it has
\begin{eqnarray}\label{enstrophy}
X^{\mathrm{enstrophy}}_{\bf k} = 1,
\end{eqnarray}
In addition to these two, there is an extra invariant
with
\begin{align}\label{TheExtraInv}
X^{\mathrm{extra}}_{\bf k}=\frac{1}{f^5 p}
\left[\arctan\frac{f(q+p\sqrt{3})}{k^2}	
\;-\; \arctan\frac{f(q-p\sqrt{3})}{k^2}
\;-\; \frac{2\sqrt{3}f p}{f^2+k^2} \right].
\end{align}
\end{subequations}
This expression is nonsingular as $f\rightarrow 0$
\begin{eqnarray}
	X^{\mathrm{extra}}_{\bf k}\simeq 8\sqrt{3}\,
    p^2\,\frac{p^2+5q^2}{5k^{10}}
	\;-\; 8\sqrt{3}\, p^2\,\frac{5p^4+42p^2q^2+21q^4}{7k^{14}}\,f^2\;
    +\;O(f^4).
\end{eqnarray}

The notion of \emph{adiabatic conservation} here is similar to that in
the theory of dynamical systems \cite{LandauLifshM}: The adiabatic
invariants are conserved \emph{approximately} over \emph{long time}.
However, here adiabatic conservation is due not to the slowness of
parameter change in time, but to the slowness of \emph{spatial} change
and to the \emph{smallness of the wave amplitudes}. This adiabatic
conservation is due to the presence of \emph{two} small parameters (see
Section \ref{Sect:SmallP}): First is the strength of nonlinearity,
compared to the beta effect [see (\ref{epsilon})], and second is the
degree of spatial inhomogeneity, i.e., the slowness of the dependence
of the Coriolis force on the latitudinal coordinate $y$, compared to
the length scale $L$ of field variation and the Rossby radius of
deformation [see (\ref{b})].

In the present paper, we derive adiabatic conservation of the above
integrals in the lowest possible non-trivial---leading---orders. We aim
here only to establish the fact of adiabatic conservation at minimal
accuracy (although the actual conservation accuracy might be higher, or
the conservation time interval might be longer). Our derivation is
formal asymptotic, but we take special care that no secular terms
appear.

In the present paper we consider the simplest possible case, making two
simplifications:
\begin{enumerate}
\item We use the beta-plane approximation (disregarding
    complications of spherical geometry).

\item We assume the fields ($u,v,h$) vanish at infinity, i.e., at
    the periphery of the beta plane. The same assumption is made
    for the quasigeostrophic equation (\ref{CHM}) when considering
    its invariants. Without this assumption, we need to account for
    the boundary terms. These can be dealt with, but their account leads to
    heavy mathematical calculations, which will not be presented
    here.
\end{enumerate}

The central result of our derivation---which allows us to establish the
extra invariant near the equator---is the possibility to cancel ``small
denominators'' at $f \rightarrow 0$; see equations (\ref{MN}),
(\ref{TMN}).

We derive the extra invariant in Sec.\ \ref{Sect:Adiabatic}, but first
we demonstrate the connection between the invariant and zonal jets.

\section{The emergence of zonal jets}
\label{Sect:Zonal}
\setcounter{equation}{0}

\subsection{Why the extra invariant implies the emergence of zonal jets}
\label{Sect:Emerge}

The approximate conservation of the energy and enstrophy, contained in
the Rossby component (see Sec.\ \ref{Sect:Extra}) implies the inverse
cascade of Rossby wave energy. At the same time (as we will see now), the
presence of the extra invariant ensures the anisotropy of the inverse
cascade: The energy is transported not just towards the origin, but
specifically to the region of the ${\bf k}=(p,q)$-plane around the
$q$-axis ($|p| \ll |q|$), which corresponds to zonal jets.

\begin{figure}
\begin{center}
\epsfxsize=\textwidth
\epsffile{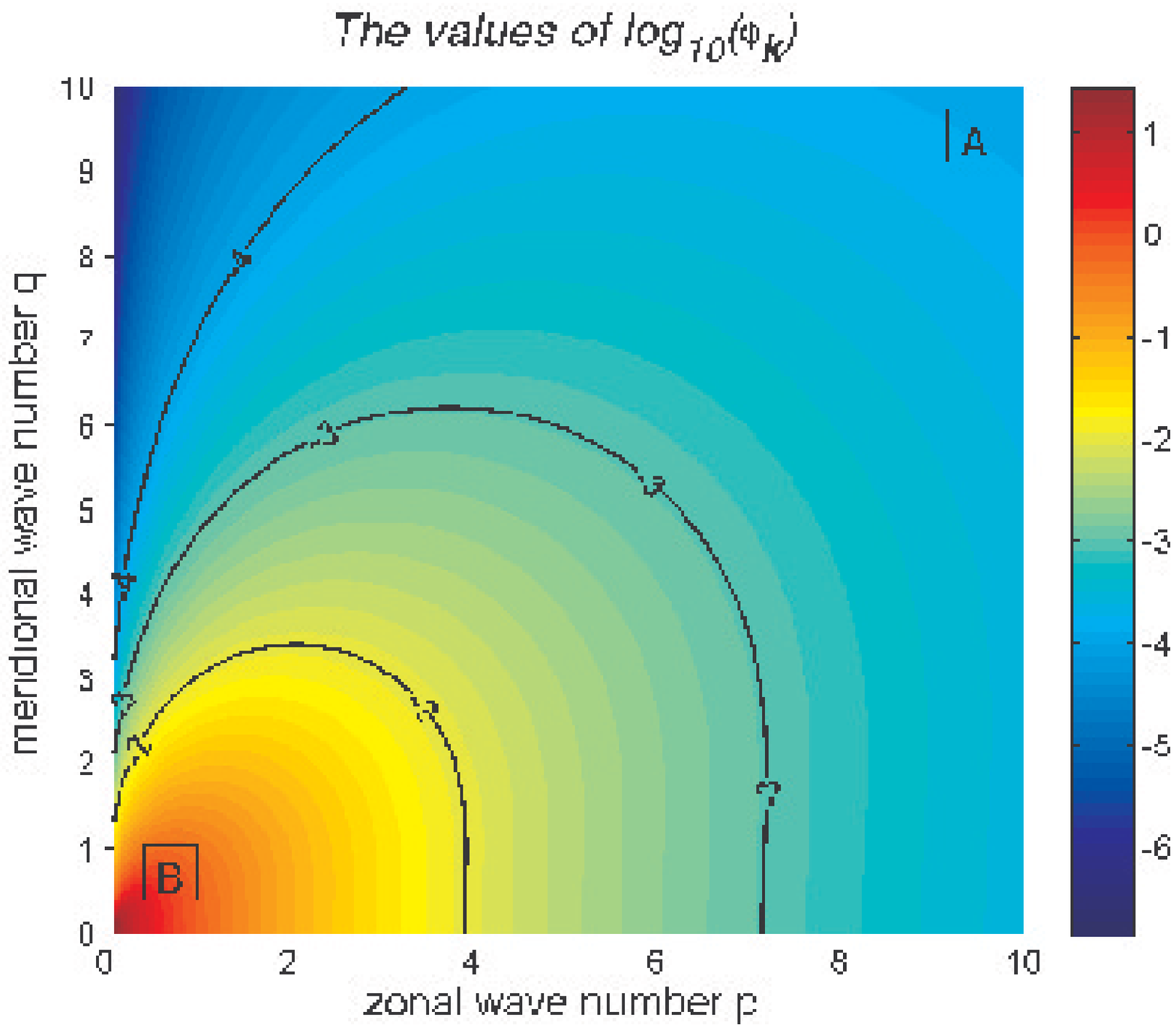}
\caption{Contour plot of $\log_{10}(\phi_{\bf k})$; the ratio
$\phi_{\bf k}$ measures how much extra invariant is carried per unit
energy by an excitation with wave vector ${\bf k}=(p,q)$; see
(\ref{TheIntegral1})--(\ref{TheExtraInv1}). The plot spans the range
$0.1 \le p \le 10, \; 0 \le q \le 10$, while $f=1$. The values of the
ratio $\phi_{\bf k}$ at the centers of boxes $A$ and $B$ differ by
roughly a factor of $7\,000$. Therefore, the only way for the energy
to transfer towards the origin (via the inverse cascade---be it local
or nonlocal) is for it to `squeeze' around the $q$-axis.}
\label{fig:BalanceArgument}
\end{center}
\end{figure}

Indeed, the extra invariant can be written in the form
\begin{eqnarray}\label{TheIntegral1}
I = \int \phi_{\bf k} \; \varepsilon_{\bf k}\; dp\,dq\,,
\end{eqnarray}
where $\varepsilon_{\bf k}$ is the Rossby wave energy spectrum, and
$\phi_{\bf k}$ is the ratio of the extra invariant spectral density to
the energy spectral density
\begin{eqnarray}\label{TheExtraInv1}
\phi_{\bf k}=\frac{f^2+k^2}{f^5\,p}
\left[\arctan\frac{f(q+p\sqrt{3})}{k^2}	
\;-\; \arctan\frac{f(q-p\sqrt{3})}{k^2}
\;-\;  \frac{2\sqrt{3}\;f p}{f^2+k^2} \right],
\end{eqnarray}
see (\ref{TheExtraInv}) and (\ref{energy}). Figure
\ref{fig:BalanceArgument} shows a contour plot of the values of the
ratio $\phi_{\bf k}$ (on a logarithmic scale) vs. ${\bf k}$.
We pose the following question: Is it possible for the energy from the
region $A$ in Fig.\ \ref{fig:BalanceArgument} be transferred (via the
inverse cascade) into the region $B$? The value of the ratio $\phi$ in
the region $B$ is about $7 \times 10^3$ times greater than its value in
the region $A$. So, if the transfer $A \rightarrow B$ did occur, the
value $I$ of the extra invariant (\ref{TheIntegral1}) would
significantly increase. The only way for the inverse cascade to
transfer the energy towards the origin would be to transport the energy
(on average) along the level lines of the function $\phi_{\bf k}$.
Thus, the dynamics must display anisotropic ``Bose condensation'':
Spatial anisotropy, which is only weakly broken in small scale
dynamics, becomes ever more strongly broken on large scales.

\subsection{Why zonal jets should be more clearly observed near the equator}
\label{Sect:Why}

\begin{figure}
\begin{center}
\epsfxsize=\textwidth
\epsffile{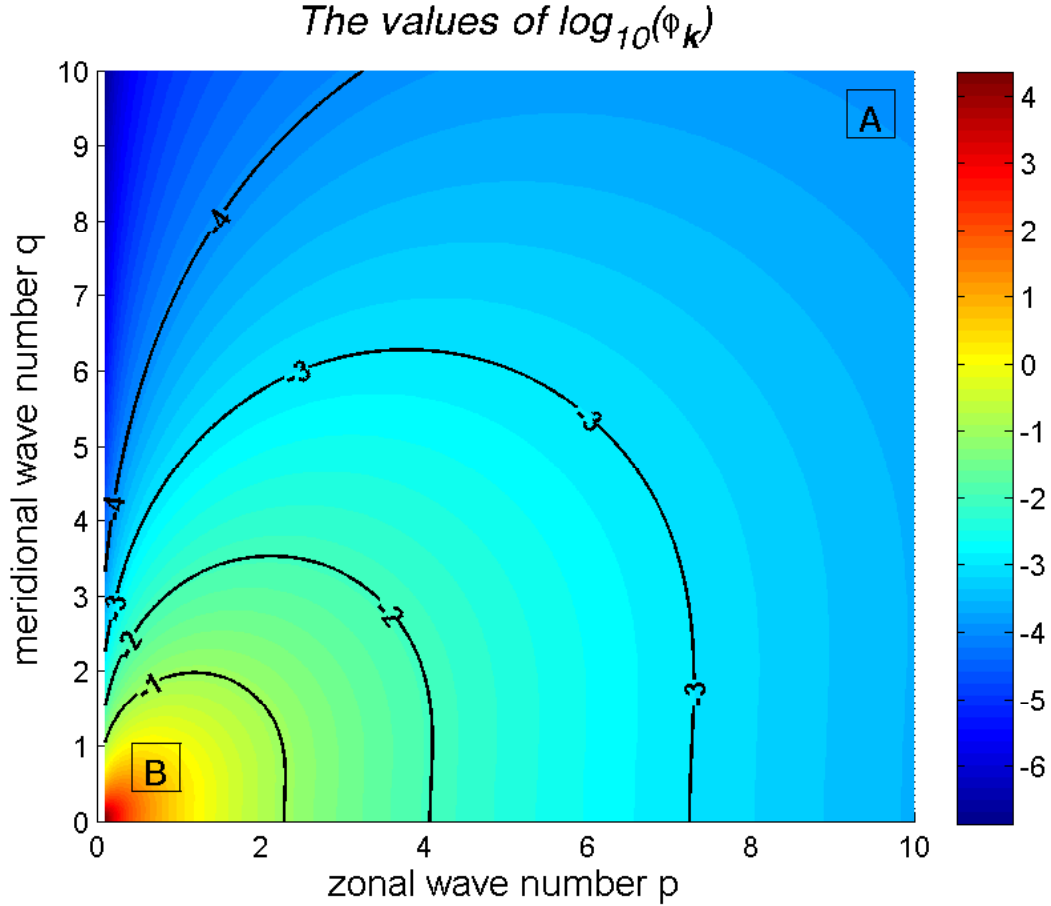}
\caption{Same as Fig.\ \ref{fig:BalanceArgument}, but for an equatorial
region, i.e., for small $f$; for this particular figure, $f=0.03$.
Comparison of Fig.\ \ref{fig:BalanceArgument} and Fig.\
\ref{fig:BalanceArgument0} demonstrates that the energy transfer towards
zonal jets should be more pronounced near the equator than at
mid-latitudes. The values of the ratio $\phi_{\bf k}$ at the centers of
boxes $A$ and $B$ now differ by roughly a factor of $30\,000$.}
\label{fig:BalanceArgument0}
\end{center}
\end{figure}

The difference between the values of $\phi$ in the regions $A$ and $B$
increases as $|f|$ decreases ($f \approx 0$ near the equator). Figure
\ref{fig:BalanceArgument0} shows the values of the ratio $\phi_{\bf k}$
when $f=0.03$.
The value of $\phi$ in the center of region $B$ is now about $3\times
10^4$ times greater than its value in region $A$. Therefore, on the
equatorial beta-plane, $f \approx 0$, the inverse cascade is forced to
transfer energy even closer to the $q$-axis.

\begin{figure}
\begin{center}
\epsfxsize=\textwidth
\epsffile{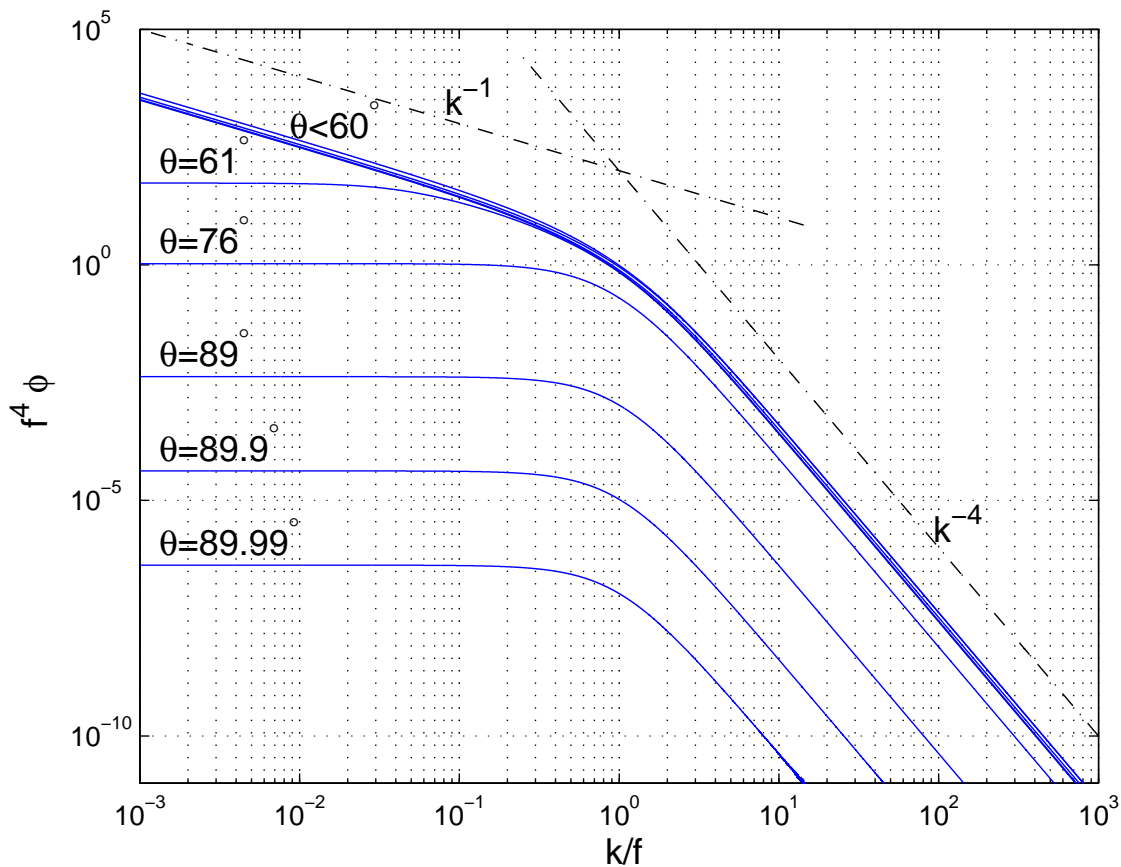}
\caption{The quantity $f^4 \phi$ is a function of $k/f$ while the polar
angle $\theta$ is held fixed. The curves correspond to the ten rays with
$\theta=0^\circ, 15^\circ, 30^\circ, 45^\circ, 60^\circ, 61^\circ,
76^\circ, 89^\circ, 89.9^\circ, 89.99^\circ$. Curves with $\theta \le
60^\circ$ (including $\theta=60^\circ$) are almost indistinguishable and
marked by one label $\theta < 60^\circ$. If $\theta > 60^\circ$, the
curves become horizontal when $k \ll f$.}
\label{fig:XpEonK}
\end{center}
\end{figure}

The more pronounced formation of zonal jets near the equator can be
seen quantitatively from Fig.\ \ref{fig:XpEonK}, which shows the
dependence (on a log-log scale) of the ratio $\phi$ vs.\ the wave
number $k$ at fixed polar angles $\theta$ (due to symmetries, we need
consider polar angles only in the range $0^\circ \leq \theta \leq
90^\circ$).
The curves shown in Fig.\ \ref{fig:XpEonK} are steeper for large $k/f$
than for small $k/f$: For large $k/f$ they vary as $k^{-4}$, while for
small $k/f$ they vary as $k^{-1}$ if $\theta \le 60^\circ$ and as $k^0$
if $\theta > 60^\circ$. Therefore, during the inverse cascade, the
ratio $\phi$ increases more significantly with decreasing $k$ if $f
\approx 0$ (near the equator).

For example, if the energy originated in the region $k/f > 20$, then
the inverse cascade must transfer this energy (on average) into the
sector $89.9^\circ < \theta < 90^\circ$. [Indeed, $f^4 \phi$ at ($k/f >
20$ and all $\theta$) is less than $f^4 \phi$ at ($\theta=89.9^\circ$ and
$k/f \rightarrow 0$).] Such `tight squeezing' of energy around the
$q$-axis hardly can be accounted for by the relative decrease of the
nonlinearity as $f \rightarrow 0$ (which might be expected in some
situations).

\subsection{Long-wavelength limit --- polar angle $60^\circ$}
\label{Sect:LongW}

\begin{figure}
\begin{center}
\epsfxsize=\textwidth
\epsffile{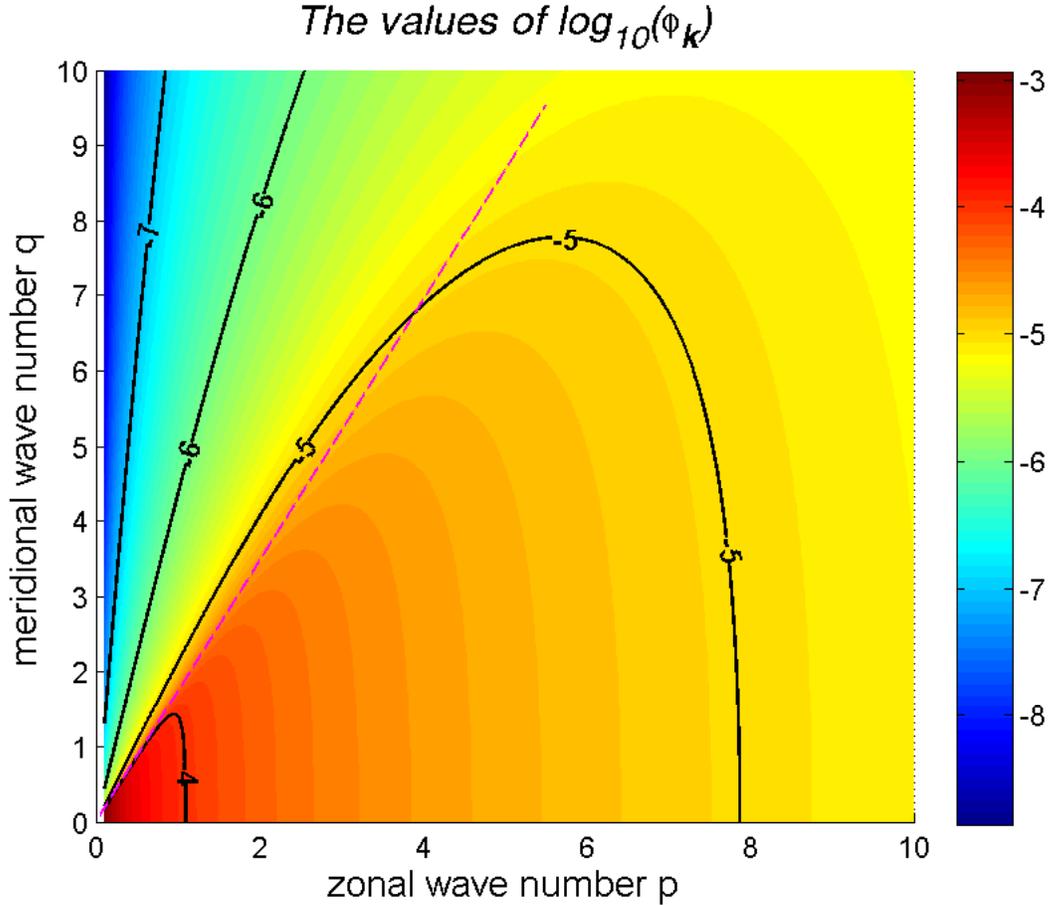}
\caption{
When $k \ll f$, the extra invariant forces energy to accumulate in the
sector $60^\circ < \theta <90^\circ$ (cf. Fig.\
\ref{fig:BalanceArgument}); for this particular figure, $f=30$. The
dashed ray marks polar angle $60^\circ$.}
\label{fig:BalanceArgument30}
\end{center}
\end{figure}

Now let us consider the opposite limit where $k/f$ is small. According
to Fig.\ \ref{fig:XpEonK}, the inverse cascade can now transfer energy
anywhere into the sector
\begin{eqnarray}\label{sector}
60^\circ < \theta < 90^\circ.
\end{eqnarray}
This is exactly the sector that was found \cite{GlWe} on the
basis of satellite altimeter observations of the spectra of very long
mid-latitude Rossby waves (with periods of several years). The sector
(\ref{sector}) is clearly visible in the contour plot of
$\log_{10}(\phi_{\bf k})$ for small $k/f$---see Fig.\
\ref{fig:BalanceArgument30}; it shows the values of the ratio
$\phi_{\bf k}$ when $f=30$. The magnitude of $\phi_{\bf k}$ drops
sharply when the polar angle $\theta$ increases beyond $60^\circ$; it
is clear that, following any level curve beginning at larger $k$, one
may approach the origin only through the sector (\ref{sector}).

We see that if the energy is generated at large scales (much greater
than the Rossby radius of deformation) then the balance argument, based
on the extra invariant, does not require the inverse cascade to
accumulate energy in zonal flows. This conclusion agrees with the
investigation \cite{OkuMasu}, which reported ``suppression of the
Rhines effect'' for large $f$.

\bigskip

To conclude this Section, we note that the existence of the extra
invariant and the balance argument (described in Secs.\
\ref{Sect:Emerge}, \ref{Sect:Why}, \ref{Sect:LongW}) holds for a wide
class of wave systems \emph{with Rossby dispersion law}. When the
\emph{nonlinearity} is taken into account, then for some special
forcing, the energy can still concentrate in zonal flows, even with
large $f$, see \cite{BaZaSimilarity}. In that paper it was also found
that, in the \emph{short wave} case (or \emph{near the equator}, Sec.\
\ref{Sect:Why}), specially arranged forcing can accelerate the
formation of zonal jets.

The presented balance argument for the emergence of zonal jets has the
appeal that it is based on a (previously unnoticed) conservation law.
However, this argument crucially relies on the assumption of weak
nonlinearity. Whether the nonlinearity is weak depends both on the
forcing strength and on the location of sources and sinks in Fourier
space. Physical examples often show that the turbulence is weak in the
large-scale part of the inertial range, in spite of the fact that the
energy spectrum becomes infinite when ${\bf k} \rightarrow 0$; e.g.,
consider sea wave turbulence \cite{ZakhSurf}. In the case of
geostrophic turbulence, the ratio of the magnitude of nonlinear terms
to the magnitude of linear terms in the quasigeostrophic equation is
the Rhines number $\epsilon=A/(\beta L^2)$. [For simplicity, we
consider here the short-wave limit, when the Rossby radius of
deformation is effectively infinite; more refined estimates will be
given in Sec.\ \ref{Sect:SmallP}.] During the inverse cascade the
length scale $L$ increases, while the velocity scale $A$ stays roughly
constant (determined by the energy), and so, $\epsilon \rightarrow 0$.

\section{Derivation of the adiabatic invariants}
\label{Sect:Adiabatic}
\setcounter{equation}{0}

In this Section, we demonstrate approximate conservation of the
quadratic invariants (\ref{TheIntegral}) with the kernels
(\ref{kernels}).

\subsection{Refining the Rossby mode}
\label{Sect:mode}

Dropping the nonlinear terms in (\ref{uvh}) leads to the linearized system
\begin{subequations}\label{Luvh}
\begin{eqnarray}
u_t &=&- f v - h_x,
\label{Lu} \\
v_t &=&f u - h_y,
\label{Lv} \\
h_t &=&-u_x - v_y,
\label{Lh}
\end{eqnarray}
\end{subequations}
while the linearized perturbational potential vorticity (\ref{PoVoL})
obeys the equation
\begin{equation}\label{LQt}
{\mathcal Q}_t=-\beta\,v.
\end{equation}

Because of the $y$-dependence of coefficients in (\ref{Luvh}), we need
to refine the Rossby mode. Let us add to ${\mathcal Q}$ a correction
${\mathcal R}$ (to be determined below) that is of higher order with
respect to the parameter $\beta$, to obtain a new field
\begin{eqnarray}\label{sQR}
s={\mathcal Q}+{\mathcal R}
\end{eqnarray}
such that in the linear approximation (\ref{Luvh}) the derivative $s_t$
will be determined by $s$ alone (not by $u,v,h$ taken separately or in
any other combination, besides $s$). Calculations show that we need to
construct ${\mathcal R}$ such that
\begin{eqnarray}\label{R}
(f^2-\Delta){\mathcal R}=\beta(f u + h_y).
\end{eqnarray}
Indeed, if ${\mathcal R}$ is determined by (\ref{R}) then
\begin{eqnarray}
(f^2-\Delta)s_t=(f^2-\Delta)({\mathcal Q}_t+{\mathcal R}_t)
=-(f^2-\Delta)(\beta v)+\beta (fu+h_y)_t,
\end{eqnarray}
and, according to the dynamics (\ref{Luvh}),
\begin{eqnarray}\label{Rossby0}
(f^2-\Delta)s_t=\beta s_x+2\beta'v_y+\beta''v.
\end{eqnarray}
Here the right hand side is $\beta s_x\;+\;O(\beta^2)$, so that the
non-$s$ terms are indeed pushed to higher order. Neglecting the higher
order terms, we see that the $s$-mode has the Rossby
wave dispersion
\begin{eqnarray}\label{Rossby}
\Omega_{\bf k}=-\frac{\beta p}{f^2+k^2}\,.
\end{eqnarray}

\subsection{Small parameters}
\label{Sect:SmallP}

The energy of the system (\ref{uvh}) is
\begin{eqnarray}\label{E}
E = \frac{1}{2}\,{\bar H} c^2 \int [(u^2+v^2)(1+h)\;+\;h^2]\;dx\,dy.
\end{eqnarray}
In the weakly nonlinear limit, the integrand of the energy (\ref{E})
reduces to $u^2+v^2+h^2$, and we assume $u,v,h$ to have the same
magnitude $A$.

As mentioned above, we exploit \emph{two small parameters}: First, the
field magnitude $A$ should be ``small'', compared to the beta-effect
[see (\ref{epsilon})]. Second, the Coriolis parameter $f(y)$ must be a
``slow'' function of $y$, so that $\beta(y) \equiv f'(y)$ is ``small''
[see (\ref{b})]. We first define the two \emph{non-dimensional} small
parameters when the field variations are characterized by a single
length scale $L$, being the same in $x$ and $y$ directions.

The magnitude of the linearized potential vorticity is given by
\begin{eqnarray}\label{l}
{\mathcal Q} \propto \frac{A}{l}, \quad\mbox{ where }
\quad \frac{1}{l}=\frac{1}{L}+f.
\end{eqnarray}
To measure the degree of nonlinearity we consider the ratio $\epsilon$
of the magnitude of the nonlinear convective terms in (\ref{Qt}) to the
magnitude of the linear term:
\begin{eqnarray}\label{epsilon}
\epsilon=\frac{A}{\beta l L}.
\end{eqnarray}
Near the equator (where $f \ll L^{-1}$), the nonlinearity degree
becomes the Rhines number: $\epsilon=A/\beta L^2$.

Small inhomogeneity means that $f$ changes little over the length scale
$L$. The change is $\Delta f \approx \beta L$. Away from the equator,
$\Delta f$ should be compared to $f$; near the equator it should be
compared to $L^{-1}$. Thus, to quantify the degree of spatial
inhomogeneity we use the parameter
\begin{eqnarray}\label{b}
b = \beta l L, \quad\mbox{ and so, } \quad A = \epsilon b.
\end{eqnarray}

In more general situations the length scales in $x$ and $y$ directions
can be different (which is especially relevant when considering zonal
jets). Moreover, the dynamics can be characterized by a wide range of
length scales, and they can change in time (they can easily change by
an order of magnitude during the inverse cascade). To account for
different situations, we will just keep track of powers of $A$ and
$\beta$ ($A \rightarrow 0, \beta \rightarrow 0$). To maintain the
condition that the field be small in comparison to the beta-effect, we
will assume the existence of a small parameter $\epsilon$, such that $A
\propto \epsilon \beta$. [In general, $\epsilon$ and $b$ will have a
more complex dependence on physical scales than (\ref{epsilon}) and
(\ref{b}).]

When there is a single length scale $L$, then the ${\mathcal
R}$-correction in (\ref{sQR}) is $O(A L^{-1}b)$ and is proportional to
$\beta$. However, in a general situation, with many length scales, we
can only guarantee that ${\mathcal R} \propto \sqrt{\beta}$. Indeed,
for states almost constant in the zonal direction ($\partial/\partial x
= 0$) and near the equator ($f = \beta y$) equation (\ref{R}) becomes
\begin{eqnarray}\label{Ry}
\beta^2 y^2 {\mathcal R} - \frac{\partial^2 {\mathcal R}}{\partial y^2} =
\beta(\beta y u + \frac{\partial h}{\partial y}),
\end{eqnarray}
which is reduced by rescaling $\tilde y=\sqrt{\beta} y$ to the form
\begin{eqnarray}\label{Ryy}
\tilde y^2 {\mathcal R}
- \frac{\partial^2 {\mathcal R}}{\partial \tilde y^2}
= \sqrt\beta(\tilde y u + \frac{\partial h}{\partial \tilde y}),
\end{eqnarray}
exhibiting explicitly the $\sqrt{\beta}$ scale of ${\mathcal R}$.

Since the difference between fields ${\mathcal Q}$ and $s$ is small
(proportional to $\sqrt{\beta}$), we replace ${\mathcal Q}$ in
(\ref{TheIntegral}) by $s$:
\begin{eqnarray}\label{TheIntegral0}
I \approx I^\star =\frac{1}{2} \int X_{\bf k} \;  s_{\bf k}\, s_{-\bf k}\; dp\,dq.
\end{eqnarray}

\subsection{Supplementing the quadratic extra invariant with cubic terms}
\label{Sect:Suppl}

Our central claim is that the increment $\Delta I^\star \equiv
I^\star(t) - I^\star(0)$ remains small over long times $t$. However,
this does not necessarily mean that $\dot I^\star$ is small:
$I^\star(t)$ can oscillate in time, similar to the behavior of
adiabatic invariants in the theory of dynamical systems. So, we use the
approach \cite{ZSch} and supplement the quadratic integral
(\ref{TheIntegral0}) with a cubic part
\begin{eqnarray}\label{SupplInv}
I^{\mbox{\scriptsize suppl}}=I^\star\;+\;I^{\mbox{\scriptsize cubic}};
\end{eqnarray}
then require $\dot I^{\mbox{\scriptsize suppl}}$ to vanish to leading
order. The general form of the cubic correction is
\begin{eqnarray}\label{GeneralCubicCorr}
&&I^{\mathrm{cubic}} =
\frac{1}{6}\int \left[ Y^{uuu}_{123}\, u_1 u_2 u_3 \,+\,
	                       Y^{vvv}_{123}\, v_1 v_2 v_3 \,+\,
                           Y^{hhh}_{123}\, h_1 h_2 h_3 \right]\, d_{123}
\nonumber \\
&&+\ \frac{1}{2}\int \left[ Y^{uuv}_{123}\, u_1 u_2 v_3 \,+\,
	                       Y^{uuh}_{123}\, u_1 u_2 h_3 \,+\,
	                       Y^{vvu}_{123}\, v_1 v_2 u_3 \,+\,
	                       Y^{vvh}_{123}\, v_1 v_2 h_3 \right.
\nonumber \\
&&\hspace{1 cm}    \left. +\ Y^{hhu}_{123}\, h_1 h_2 u_3 \,+\,
	                       Y^{hhv}_{123}\, h_1 h_2 v_3 \right]\, d_{123}
\;+\;              \int        Y^{uvh}_{123}\, u_1 v_2 h_3\,  d_{123}
\end{eqnarray}
with 10 kernels $Y^{uuu}, Y^{vvv}, \ldots$. Here and throughout the
rest of this paper a subscript $j$ stands for the wave vector ${\bf
k}_j=(p_j,q_j)$ ($j=1,2,3$); e.g. $u_1=u_{{\bf k}_1}$, likewise,
$Y_{123}=Y({\bf k}_1,{\bf k}_2,{\bf k}_3)$ for any kernel $Y$, and
$\delta_{123}=\delta({\bf k}_1+{\bf k}_2+{\bf k}_3)$, $d_{123}=d{\bf
k}_1 \, d{\bf k}_2 \, d{\bf k}_3$. In addition, a subscript $-j$ will
denote $-{\bf k}_j$, in particular, $Y_{-123}=Y(-{\bf k}_1,{\bf
k}_2,{\bf k}_3)$.

The form of the shallow water system allows us to consider the
following more simple form of the cubic correction
\begin{eqnarray}\label{CubicCorr}
I^{\mbox{\scriptsize cubic}}=\frac{1}{2}\int s_1\, s_2\,
\left[ M_{123}\, u_3 \,+\, N_{123}\, v_3
\,+\, T_{123}\, h_3 \right] \, d_{123}
+ \frac{1}{6}\int Y_{123}\, s_1 s_2 s_3\, d_{123}
\end{eqnarray}
with only 4 kernels $M, N, T,$ and $Y$ instead of the ten kernels in
(\ref{GeneralCubicCorr}). The general form (\ref{GeneralCubicCorr}) and
the simplified form (\ref{CubicCorr}) lead to the same final result. A
much longer calculation demonstrates that the kernels in
(\ref{GeneralCubicCorr}) must be related in such a way that the terms
may be collected in the form (\ref{CubicCorr}).

When calculating $\dot I^{\mbox{\scriptsize suppl}}$,
we will have contributions of different nonlinearity orders
\begin{subequations}\label{dIdt0}
\begin{eqnarray}
\dot I^{\mbox{\scriptsize suppl}}&=&
\dot I^\star \quad\mbox{due to the {\it linear} terms in the equations}
\label{QuadrLin}\\
&+&\dot I^\star \quad\mbox{due to the {\it quadratic} terms in the equations}
\label{QuadrQuadr}\\
&+&\dot I^{\mbox{\scriptsize cubic}} \quad\mbox{due to the {\it linear} terms in
the equations}\label{CubicLin}\\
&+&\dot I^{\mbox{\scriptsize cubic}} \quad\mbox{due to the {\it quadratic} terms
in the equations}\label{CubicQuadr}
\end{eqnarray}
\end{subequations}
We will see that the first contribution (\ref{QuadrLin}) vanishes
automatically. Our goal is to show that it is possible to find the
cubic correction $I^{\mbox{\scriptsize cubic}}$---with non-singular
(uniformly bounded) kernels---such that the next two contributions
(\ref{QuadrQuadr}) and (\ref{CubicLin}) exactly cancel each other,
implying that, indeed, $\dot I^{\mbox{\scriptsize suppl}}$ is
determined by only higher order terms (\ref{CubicQuadr}). Formally, one
can always achieve such cancellation for any wave system, but the
corresponding kernels will generally be singular. The possibility to
escape these singularities takes place only for \emph{very few} systems
\cite{ZSch0,ZSch}. Significantly, our results demonstrate that the
rotating shallow water system is among them.

\subsection{The time derivative $\dot I^\mathrm{suppl}$}
\label{Sect:dotI}

We can always assume the obvious symmetries:
\begin{eqnarray}
X_{{\bf k}} = X_{-{\bf k}}, \;\; M_{123} = M_{213},
\;\; N_{123} = N_{213},\;\; T_{123} = T_{213},
\;\; Y_{123} = Y_{213} = Y_{321}.
\label{TransposeSymm}
\end{eqnarray}

According to the rotating shallow water dynamics (\ref{uvh}),
(\ref{Qt}), along with the definition (\ref{sQR}), we have to leading
orders
\begin{eqnarray}\label{dIdt}
\dot I^\mathrm{suppl} &&= \int X_1\, s_{-1}\,
\left[-i\Omega_1\,s_1+\int(-ip_1u_3-iq_1v_3)\,s_2\,
\delta_{-123}\,d_{23}\right]\,d_1
\nonumber\\
&&+\ \frac{1}{2}\int s_1\,s_2\,\left[M_{123}(f v_3 - ip_3 h_3)\,+\,
N_{123}(-f u_3-iq_3 h_3) \right.
\nonumber \\
&&\hspace{5cm}+\ \left. T_{123}(-ip_3 u_3 - iq_3v_3)\right]\, d_{123}
\nonumber \\
&&+ \frac{1}{2} \int Y_{123}\, s_1 \,s_2\,
(-i\Omega_3\, s_3) \,d_{123}
\end{eqnarray}
[in accordance with our notations (introduced in Sec.
\ref{Sect:Suppl}), $\delta_{-123}=\delta(-{\bf k}_1+{\bf k}_2+{\bf
k}_3)$]. The equation (\ref{dIdt}) explicitly displays the
contributions summarized in (\ref{dIdt0}\emph{abc}). We will determine
the kernels $X,M,N,T,Y$ from the requirement that the right hand side
of this expression vanish, and that they be nonsingular.

First, the integral $\int X_1\,\Omega_1\, s_{-1}s_1\,d_1$ vanishes
automatically since $X_{\bf k}$ is even, and $\Omega_{\bf k}$ is odd.

Using (\ref{PoVoL}) and (\ref{sQR}), we substitute
\begin{eqnarray}
u_3=\frac{ip_3 v_3 - f h_3 - s_3}{iq_3}
\end{eqnarray}
into (\ref{dIdt}) and collect terms into three groups: those containing
(1) $ssv$, (2) $ssh$, and (3) $sss$:
\begin{eqnarray}\label{dIdt1}
&&\dot I^\mathrm{suppl} = \frac{1}{2}\int\,d_{123}\, s_1\, s_2
\nonumber\\
&&\times\ \left\{\frac{v_3}{q_3}\left[f q_3 M_{123}
-f p_3 N_{123}-k_3^2 T_{123}+
i(p_3 p_1 X_1+p_3 p_2 X_2
+ q_3 q_1 X_1+q_3 q_2 X_2)\delta_{123}\right]\right.
\nonumber\\
&&\ \ \ \ +\ \frac{h_3}{q_3}\left[i p_3 q_3 M_{123}p_3q_3
+i(f^2+q_3^2)N_{123} - f p_3 T_{123}
+ f(p_1X_1+p_2X_2)\delta_{123}\right]
\nonumber\\
&&\ \ \ \ \left.+\ \frac{s_3}{q_3}\left[if N_{123}-p_3T_{123}
+ (p_1X_1+p_2X_2)\delta_{123}\right] \right\}
\nonumber\\
&& \hspace*{2cm}  -\frac{i}{6}\int Y_{123}\,s_1\,s_2\,s_3\,
(\Omega_1+\Omega_2+\Omega_3) \,d_{123}
\end{eqnarray}

\subsection{Canceling ``small denominators'' which are due to the
equatorial limit ($f \ll k$)} \label{Sect:SmDeEq}

Equating to zero the coefficients of $v_3$ and $h_3$ produces a system
of two linear algebraic equations, which we solve for the kernels $M$
and $N$:
\begin{subequations}\label{MN}
\begin{eqnarray}
M_{123}&=&\frac{i q_3 T_{123}}{f}-\label{kernelM}
\\
&&\frac{i q_3\left[(p_1p_3+q_1q_3)X_1+(p_2p_3+q_2q_3)X_2\right]
+i f^2(q_1X_1+q_2X_2)}{f(f^2+k_3^2)}\;\delta_{123},
\nonumber \\
N_{123}&=&-\frac{i p_3 T_{123}}{f}+\label{kernelN}
\\
&&\frac{i p_3\left[(p_1p_3+q_1q_3)X_1+(p_2p_3+q_2q_3)X_2\right]
+i f^2(p_1X_1+p_2X_2)}{f(f^2+k_3^2)}\;\delta_{123}
\nonumber
\end{eqnarray}
\end{subequations}
These expressions have an apparent singularity when $f\rightarrow 0$,
which would invalidate our perturbational expansion (e.g., making the
cubic correction larger than the main quadratic part). However, it is
possible to choose the kernel $T$ in such a way as to eliminate the
singularities in $M$ and $N$. \emph{Both} expressions (\ref{MN}) become
non-singular as $f\rightarrow 0$ if we take
\begin{subequations}\label{TMN}
\begin{eqnarray}
T_{123}&=&\frac{p_3(p_1X_1+p_2X_2)+q_3(q_1X_1+q_2X_2)}
{f^2+k_3^2}\;\delta_{123}.\label{T}
\end{eqnarray}
This cancels all terms proportional to $1/f$, and produces
\begin{eqnarray}
M_{123}&=&-i\frac{f(q_1X_1+q_2X_2)}{f^2+k_3^2}\;\delta_{123}\label{M},\\
N_{123}&=& i\frac{f(p_1X_1+p_2X_2)}{f^2+k_3^2}\;\delta_{123}\label{N}.	
\end{eqnarray}
\end{subequations}
The denominators in (\ref{TMN}) still appear to be singular when $f
\rightarrow 0,\; k_3\rightarrow 0$ simultaneously. However, due to the
presence of the delta functions $\delta_{123} \equiv \delta({\bf
k}_1+{\bf k}_2+{\bf k}_3)$, the condition $k_3 \rightarrow 0$ implies
${\bf k}_1+{\bf k}_2 \rightarrow 0$, and so, the expressions
$p_1X_1+p_2X_2$ and $q_1X_1+q_2X_2$ in the numerators (\ref{TMN}) are
linear in $p_3,q_3$ when $k_3\rightarrow0$. Therefore, the numerators
in (\ref{TMN}) are quadratic in $f, p_3, q_3$, and the expressions
(\ref{TMN}) are bounded.

Substituting (\ref{TMN}) into (\ref{dIdt1}), we find
\begin{eqnarray}\label{dIdt2}
\dot I^\mathrm{suppl} &=& \frac{1}{2}\int\,
\left[\frac{p_3q_1-p_1q_3}{f^2+k_3^2}X_1\,+\,
\frac{p_3q_2-p_2q_3}{f^2+k_3^2}X_2\right]
\,s_1\,s_2\,s_3\, \delta_{123}\,d_{123}
\nonumber\\
&-&\frac{i}{6}\int Y_{123}\,s_1\,s_2\,s_3\,
(\Omega_1+\Omega_2+\Omega_3)\,d_{123}
\end{eqnarray}

\subsection{Canceling the triad resonance ``small denominators''}
\label{SmallDenominators}

Note the obvious identity
\begin{eqnarray}
\left|\begin{array}{cc}  p_1 & p_2 \\ q_1 & q_2 \end{array}\right|
=\left|\begin{array}{cc} p_2 & p_3 \\ q_2 & q_3 \end{array}\right|
=\left|\begin{array}{cc} p_3 & p_1 \\ q_3 & q_1 \end{array}\right|
\mbox{ when } {\bf k}_1+{\bf k}_2+{\bf k}_3 = 0
\label{|.|}
\end{eqnarray}
(to see, e.g., the first equality, substitute $p_1=-p_2-p_3, \;
q_1=-q_2-q_3$ from the last equation). Because of (\ref{|.|}), equation
(\ref{dIdt2}) reduces to
\begin{eqnarray}\label{dIdt3}
\dot I^\mathrm{suppl} &=& \frac{1}{2}\int\,
\frac{p_3q_1-p_1q_3}{f^2+k_3^2}(X_1-X_2)
\,s_1\,s_2\,s_3\, \delta_{123}\,d_{123}\nonumber\\
&-&\frac{i}{6}\int Y_{123}\,s_1\,s_2\,s_3\,
(\Omega_1+\Omega_2+\Omega_3)\,d_{123}.
\end{eqnarray}
Symmetrizing the first term on the right hand side (over all
permutations of the indices 1,2,3), and using again the identity
(\ref{|.|}), we see that (\ref{dIdt3}) vanishes if
\begin{equation}\label{Y}
Y_{123} = \frac{p_1 q_2 - p_2 q_1}
{i(\Omega_1 + \Omega_2 + \Omega_3)}
\;\delta_{123}
\left[\frac{ X_2 -  X_1}{f^2 + k_3^2}
+ \frac{ X_3 -  X_2}{f^2 + k_1^2}
+ \frac{ X_1 -  X_3}{f^2 + k_2^2} \right].
\end{equation}
It is apparent that this expression is singular at the points $({\bf
k}_1, {\bf k}_2, {\bf k}_3)$ satisfying the resonance relations
\begin{subequations}\label{Resonance}
\begin{eqnarray}
{\bf k}_1 + {\bf k}_2 + {\bf k}_3 = 0,
\label{ResonanceSpace} \\
\Omega_{{\bf k}_1}+\Omega_{{\bf k}_2}+\Omega_{{\bf k}_3}=0,
\label{ResonanceTime}
\end{eqnarray}
\end{subequations}
\emph{unless} the expression in square brackets vanishes at these
points. We will see now that the latter is indeed the case.

On the resonance manifold (\ref{Resonance}), the bracketed expression
in (\ref{Y}) may be put in the form
\begin{equation}
[\ldots] = \frac{p_1 \Omega_2 -  p_2 \Omega_1} {\beta\, p_1 p_2 p_3}
\left(p_1 X_1 + p_2  X_2 + p_3  X_3 \right).
\end{equation}
To obtain this, the dispersion relation (\ref{Rossby}) must be used
along with the identities
\begin{equation}
\left|\begin{array}{cc}  p_1 & p_2 \\ \Omega_1 & \Omega_2 \end{array}\right|
=\left|\begin{array}{cc} p_2 & p_3 \\ \Omega_2 & \Omega_3 \end{array}\right|
=\left|\begin{array}{cc} p_3 & p_1 \\ \Omega_3 & \Omega_1 \end{array}\right|
\mbox{ when}\left\{\begin{array}{c}
p_1 + p_2 + p_3 = 0\\
\Omega_1 + \Omega_2 + \Omega_3 = 0,
\end{array}\right.
\end{equation}
which are similar to (\ref{|.|}). Thus, to obtain a non-singular form
for $Y_{123}$, we have to require that the function $p X_{\bf k}$ be
conserved in the triad resonance interactions, i.e., that the equation
\begin{equation}
p_1  X_1 + p_2  X_2 + p_3  X_3 = 0
\label{WebG}
\end{equation}
must hold in all points $({\bf k}_1, {\bf k}_2, {\bf k}_3)$ of the
resonance manifold (\ref{Resonance}).

\subsection{Kernel $X$}

The requirement (\ref{WebG}) is satisfied for the following five
functions:
\begin{subequations}\label{choices}
\begin{eqnarray}
p X_{\bf k} &=& \Omega_{\bf k},\label{En}\\
p X_{\bf k} &=& p,             \label{p}\\
p X_{\bf k} &=& q,             \label{q} \\
p X_{\bf k} &=& \xi_{\bf k}\stackrel{\mathrm{def}}{=}
\frac{1}{2}\ln\left[f^2(q+p\sqrt{3})^2+k^4\right] -
\frac{1}{2}\ln\left[f^2(q-p\sqrt{3})^2+k^4\right]
\label{xi} \\
p X_{\bf k} &=& \eta_{\bf k}\stackrel{\mathrm{def}}{=}
\arctan\left[f (q+p\sqrt{3})/k^2\right] -
\arctan\left[f (q-p\sqrt{3})/k^2\right]. \label{eta}
\end{eqnarray}
\end{subequations}
For the functions $\xi_{\bf k}$ and $\eta_{\bf k}$ see \cite{B1991}; their physical meaning remains unclear, let alone their possible relation to some continuous symmetries.

For the function (\ref{En}), the integral $I$ is the \emph{energy} of
the Rossby component.

The function (\ref{p}) corresponds to the \emph{enstrophy}. More
precisely, in this case the integral $I$ is the zonal (East-West)
momentum, which is a linear combination of the energy and enstrophy.

The function (\ref{q}) corresponds to the North-South momentum
\cite[][]{LoHGill}. However, this choice fails to give a physically
meaningful quantity in real (coordinate) space because the
corresponding function $X_{\bf k}$ is singular (when $p \rightarrow
0)$. This singularity means that respective invariant in real space
\begin{eqnarray}
I = \frac{1}{2} \int X({\bf r}_1,{\bf r}_2)
s({\bf r}_1,t) s({\bf r}_2,t) d{\bf r}_1 d{\bf r}_2
\label{ExtraInvariant}
\end{eqnarray}
has kernel $X({\bf r}_1,{\bf r}_2)$ which does not vanish at large
separation ${\bf r}_1-{\bf r}_2$; see \cite{BaYo} for a detailed
discussion.

The function (\ref{xi}) fails to produce an invariant either. This is
because $\xi_{\bf k}$ is even in ${\bf k}$, and so, $X_{\bf k}$ is odd,
contradicting the symmetry (\ref{TransposeSymm}).

Unlike $\xi_{\bf k}$, the function $\eta_{\bf k}$ is odd,
and the corresponding kernel $X_{\bf k}$ determines an extra invariant
for rotating shallow water dynamics. The previously described kernel
(\ref{TheExtraInv}) is a linear combination of the functions (\ref{En})
and (\ref{eta}).

The proof of the fact that the functions $\xi_{\bf k}$ and $\eta_{\bf
k}$ are conserved in triad resonance interactions has recently been
significantly simplified. The new proof is more straightforward and can
be accomplished with the aid of symbolic algebra software. Indeed,
\begin{eqnarray}
\xi_{\bf k}+i\eta_{\bf k}=\ln Z_{\bf k} \quad \mbox{where}\quad
Z_{\bf k}=\frac{if(q+p\sqrt{3})+k^2}{if(q-p\sqrt{3})+k^2},
\end{eqnarray}
($\ln$ denotes the principal branch of the complex logarithm, with
argument between $-\pi$ and $\pi$), and the required conservation
equation
\begin{eqnarray}\label{XiEta}
(\xi_1+i\eta_1)+(\xi_2+i\eta_2)+(\xi_3+i\eta_3)=0
\end{eqnarray}
implies
\begin{eqnarray}\label{Zeq}
Z_1\,Z_2\,Z_3=1.
\end{eqnarray}
Now, using (\ref{ResonanceSpace}), substitute $p_3=-p_1-p_2, \;
q_3=-q_1-q_2$ into (\ref{ResonanceTime}) and (\ref{Zeq}). These
equations may then be reduced to two polynomial equations of degree 5
in $p_1,q_1,p_2,q_2$. It is easy to check (e.g., with {\sc Mathematica}
software) that these two polynomials are identical up to a constant
factor. It follows immediately that the resonance equations
(\ref{Resonance}) imply (\ref{Zeq}), and hence that
\begin{eqnarray}\label{lnZeq}
\ln Z_1\,+\,\ln Z_2\,+\,\ln Z_3\,=\,2\pi m i,
\quad\mbox{where}\quad m=0,\pm 1, \pm 2, \ldots.
\end{eqnarray}
Continuity considerations require $m=0$ \cite{B1991}, and the
conservation (\ref{XiEta}) then follows.

Thus, there are three invariants:

\begin{itemize}

\item \emph{the energy of the Rossby component} [corresponding to
    (\ref{En})],

\item \emph{the enstrophy} [corresponding to (\ref{p})],

\item \emph{the extra invariant} [corresponding to (\ref{eta})].

\end{itemize}

\subsection{Dropping cubic terms}

The cubic terms $I^{\mbox{\scriptsize cubic}}$ have served their
purpose in the proof, and can now be dropped, similar to the argument
\cite{BaVan} for the quasigeostrophic equation. To see this, first,
note that the $\beta^2$-terms in (\ref{Rossby0}) can be neglected over
a time interval of length at most of order $\beta^{-\nu}$ with $\nu <
2$. We also need to consider time intervals containing many wave
periods, and so $\nu > 1$. For specificity, we choose $\nu = 3/2$.

Considering (\ref{dIdt}), we have neglected terms $\;\;\propto
A^3\beta^{1/2} \propto A^2 \epsilon \beta^{3/2}$ (such terms come from
neglecting ${\mathcal R}$-correction in the nonlinear terms of the
shallow water equations). Therefore, over a time $t \propto
\beta^{-3/2}$, the error can accumulate at most up to a total error
$\propto A^2\epsilon$. The $M, N,$ and $T$-corrections in
(\ref{CubicCorr}) have the order $A^3\propto A^2\epsilon\beta$; the
$Y$-correction in (\ref{CubicCorr}) has the order $A^3/\beta \propto
A^2\epsilon$ [the kernel $Y$ is proportional to $1/\beta$, while the
kernels $M, N, T$ are $O(1), \; \beta\rightarrow0$]. So, all  cubic
corrections  are within the total conservation error $\propto
A^2\epsilon$ and can be safely dropped. As alluded to earlier, these
corrections were needed in the derivation only to control oscillatory
terms; their amplitude is now seen to be small, but their time
derivative is large (has lower order).

\section{Remarks}
\label{Sect:Remarks}
\setcounter{equation}{0}
\setcounter{subsubsection}{0}

\subsection{\emph{Unique} invariant}
\label{Sect:Unique}

The existence of an extra invariant motivates a natural question: Do
there exist other invariants in the shallow water system? The answer
appears to be ``No'', although rigorous investigation of this question
has not been attempted. To elaborate, if such an invariant did exist,
then the resonance triad interaction (\ref{Resonance}) would seem to
have another conserved quantity, besides (\ref{choices}). The latter,
however, is known to be untrue \cite{BaFe}. This was established by the
connection \cite{Ferapontov} between invariants of wave interactions
and {\it Web geometry} \cite{Bl1}. It has not been ruled out, however, that
the shallow water system (\ref{RSW}) has several invariants, which
collapse into a single invariant for the quasigeostrophic equation
(\ref{CHM}); though this seems unlikely.

The connection to the Web geometry also shows that the dispersion laws
that admit extra invariants are extremely rare.
We are aware of only one other physical system (besides Rossby waves)
that possesses extra invariants. This is the generalization of the
Korteweg-de Vries (KdV) equation for two spatial dimensions
\begin{align}\label{KP}
	(\psi_t+\psi\psi_x+\psi_{xxx})_x=\psi_{yy};
\end{align}
it has dispersion law
\begin{align}
	\Omega(p,q)=-p^3-\frac{q^2}{p}.
\end{align}
Equation (\ref{KP}) is integrable via the inverse scattering method and
has infinitely many extra invariants \cite{ZSch0}. The system
(\ref{KP}) is called the Kadomtsev-Petviashvili equation of the first
kind (KP1); the Kadomtsev-Petviashvili equation of the second kind
(KP2) has a minus sign in front of the term on the right of (\ref{KP});
because of this, triad resonances do not exist at all
for KP2.

Unlike to the KP1 case, the Rossby wave triad resonance admits only one extra
invariant; and moreover, it is impossible to extend this invariant to
the next nonlinearity order \cite{BaVan}. So, the extra invariant of
the shallow water dynamics (\ref{RSW}) is an attribute of weak
nonlinearity.

The triad resonances that admit finite number of extra invariants are even more rare than the ones with infinitely many invariants (see \cite{BaFe}): The former constitute a several parameter family among all functions depending on two variables; and moreover, most of the members of this family are not even elementary functions and hardly can be dispersion laws of physical
systems.

\subsection{The impact of the extra invariant on statistical equilibrium}
\label{Sect:StatEquil}

The existence of the extra invariant may provide barriers to
statistical equilibration. The equilibrium theory for the
quasigeostrophic (\cite{W06} and references therein) and shallow water
systems \cite[][]{WP01} were derived by enforcing only the exact
conservation laws (energy, momentum, and the potential vorticity
hierarchy). Since the latter fully define the equilibrium state (under
the ergodic hypothesis), the adiabatic conservation laws will generally
be violated. Given that true equilibration is an infinite time
property, the presence of an adiabatic invariant does not lead to any
mathematical contradiction here. However, there are practical issues
since the extra invariant could greatly increase the equilibration time
scale. This issue needs to be investigated.

We should also note a parallel between the existence of the extra
invariant, determined only by the Rossby component, and the equilibrium
theory. In the latter it is found that although the inertia-gravity
waves do remove some of the initial energy to small scale surface
ripples, they do not inhibit the inverse cascade of the remaining
energy to form large-scale vortex equilibria.

\subsection{Using perturbational potential vorticity instead of its
linearization}
\label{Sect:nonlPertPoVo}

Since the extra conservation holds only in the weakly nonlinear limit,
to the same accuracy we are free to write the invariants in terms of
the perturbational potential vorticity
\begin{eqnarray}\label{PoVo}
\tilde Q = \frac{v_x-u_y+f(y)}{H}\;-\;\frac{f(y)}{\bar H}
\end{eqnarray}
instead of the \emph{linearized} perturbational potential vorticity
${\mathcal Q}$, equation (\ref{PoVoL}). We have $\tilde Q \approx \bar
H {\mathcal Q}$ (with the error due to nonlinear terms), and instead of
(\ref{TheIntegral}),
\begin{eqnarray}\label{TheIntegral2}
I = \frac{1}{2{\bar H}^2}\int X_{\bf k} \;
\tilde Q_{\bf k}\, \tilde Q_{-\bf k}\; dp\,dq,
\end{eqnarray}
where $\tilde Q_{\bf k}$ is the Fourier transform of the field $\tilde
Q$.

\subsection{Can rotating shallow water dynamics be approximated by
a single equation?}
\label{Sect:single}

There is a question whether the shallow water system (\ref{RSW}) can be
approximated \emph{near the equator} by a single equation. Certainly,
in the rigid lid approximation ($H=\mbox{const}$) the system
(\ref{RSW}) is reduced to the equation of 2D hydrodynamics with
beta-effect. However, the shallow water dynamics contain three
independent variables $u,v,H$, and accordingly the system (\ref{RSW})
contains 3 time derivatives. We allow significant deviations of $H$
from its average value $\bar H$.

We have attempted to approximate the equatorial shallow water dynamics
by a single equation
\begin{align}
	{\dot a}_1=\Omega_1 a_1 + \int W_{-123} \; a_2\, a_3\; \delta_{-123}\; d_{23}
\end{align}
for the Fourier transform $a_{\bf k}(t)$ of the stream function or some
other variable (the notation is defined in Sec. \ref{Sect:Suppl}).
However, we found that such an equation would have insufficient
accuracy to establish the extra conservation. More specifically, the
formula for the kernel $W$ would lack one more cancellation in equations
similar to (\ref{TMN}) [numerator in $W$, instead of being quadratic,
would be linear in $f, p_3, q_3$], and so, the kernel $W$ would be
singular.

\subsection{Possible fast dependence on the $y$-coordinate}
\label{Sect:fast y}

The extra conservation holds if the coefficients in the shallow water
system (\ref{uvh}) additionally contain fast, but small amplitude,
dependence on the $y$-coordinate. Such inhomogeneity may be considered
at lowest order as a resonant triad interaction between two Rossby
waves, with dispersion law (\ref{Rossby}), and one inhomogeneity wave,
with zero dispersion law:
\begin{subequations}
\label{InhomRes}
\begin{eqnarray}
p_1 &=& p_2 + p_3,
\label{pp0} \\
q_1 &=& q_2 + q_3
\label{qqq}, \\
\Omega(p_1,q_1) &=& \Omega(p_2,q_2) + 0;
\label{OmegaOmega0}
\end{eqnarray}
here ${\bf k}_1 = (p_1,q_1)$ and ${\bf k}_2 = (p_2,q_2)$ are the Rossby
wave vectors, and ${\bf k}_3 = (p_3,q_3)$ is the inhomogeneity wave
vector.

If translation symmetry is still maintained in the $x$-coordinate, one
has $p_3 \equiv 0$. For this case, one can readily see that an
arbitrary function $\varphi(p,q)$ that is even in $q$ satisfies
\begin{eqnarray}
\varphi(p_1,q_1) = \varphi(p_2,q_2) + 0
\label{phi_phi_0}
\end{eqnarray}
\end{subequations}
at each point of the resonance manifold (\ref{InhomRes}abc). Indeed,
(\ref{pp0}), with $p_3\equiv 0$, and (\ref{OmegaOmega0}) imply
$p_1=p_2$ and $|q_1|=|q_2|$. In particular, the function
(\ref{TheExtraInv}) is even in $q$, and the conservation
(\ref{phi_phi_0}) holds for $\varphi \equiv \eta$. Thus, the function
(\ref{TheExtraInv}) is conserved in triad resonant interactions of
Rossby waves with the inhomogeneity waves.

Actually, the function (\ref{TheExtraInv}) is conserved in resonant
interactions of any order $n\; (n\ge 3)$, which involve 2 Rossby waves
and $n-2$ inhomogeneity waves:
\begin{subequations}
\begin{eqnarray}
p_1 &=& p_2 + \underbrace{0+0+\ldots+0}_{n-2},
\label{pp000} \\
q_1 &=& q_2 + q_3 + \ldots+q_n
\label{qqqqq}, \\
\Omega(p_1,q_1) &=& \Omega(p_2,q_2) + \underbrace{0+0+\ldots+0}_{n-2};
\label{OmegaOmega000}
\end{eqnarray}
Indeed (\ref{pp000}) and (\ref{OmegaOmega000}) imply $p_1=p_2$ and
$|q_1|=|q_2|$, and therefore,
\begin{eqnarray}
\varphi(p_1,q_1) = \varphi(p_2,q_2) + \underbrace{0+0+\ldots+0}_{n-2}
\label{phi_phi_000}
\end{eqnarray}
\end{subequations}
for any function $\varphi(p,q)$ which is even in $q$.

\section{Conclusion}
\label{Sect:Conclusion} \setcounter{equation}{0}

The Rossby waves have been known \cite{BNZ,B1991} to possess a
rare property:  Their triad resonance admits an extra conserved
quantity:
\begin{eqnarray}\label{imply}
\left.\begin{array}{l}
{\bf k}_1 + {\bf k}_2 + {\bf k}_3 = 0,\\
\Omega({\bf k}_1)+\Omega({\bf k}_2)+\Omega({\bf k}_3)=0
\end{array}\right\}\quad\Rightarrow\quad
\eta({\bf k}_1)+\eta({\bf k}_2)+\eta({\bf k}_3)=0
\end{eqnarray}
where
\begin{align}
{\bf k}=(p,q) \quad (k^2=p^2+q^2),
\quad\quad \Omega({\bf k})=\frac{\beta p}{f^2+k^2}\;\,,\\
\eta({\bf k})=\arctan\left(f \frac{q+p\sqrt{3}}{k^2}\right)
              - \arctan\left(f \frac{q-p\sqrt{3}}{k^2}\right)\,.
\end{align}
Despite of the implication (\ref{imply}), the extra invariant $I$ is
actually independent of the energy and momentum (enstrophy) because the integrals
(\ref{TheIntegral}) or (\ref{TheIntegral1}) contain time-dependent
functions ${\mathcal Q}_{\bf k}(t)$ or $\varepsilon_{\bf k}(t)$.
[Recall that $X_{\bf k}=\eta({\bf k})/p,\; \phi_{\bf k}=\eta({\bf
k})/\Omega({\bf k})$.]

In the present paper, we have established two key results:
\begin{itemize}

\item The Rossby wave extra invariant can be extended to the
    shallow water dynamics in spite of the presence of
    inertia-gravi\textsc{}ty waves and in spite of the explicit
    inhomogeneity (the $y$-dependence of the Coriolis parameter
    $f$).

\item The shallow water dynamics possesses an extra invariant in
    the equatorial limit (when $f \rightarrow 0$, but the
    derivative $f'$ stays away from zero). This limit also leads to
    small denominators, but different from those related to the
    triad resonance. We have shown that it is possible to cancel
    these small denominators.

\end{itemize}
We have also found that for weakly nonlinear shallow water dynamics,
the presence of the extra invariant constrains the inverse cascade
energy transfer to be from small scale eddies to large scale zonal
flow. The results are also in agreement with some more specific
experimental features: more pronounced zonal jets near the equator,
when $f \rightarrow 0$, and suppression of zonal jets and the
$60^\circ$ polar angle in the energy spectrum when $f \rightarrow
\infty$ (see Sec.\ \ref{Sect:Zonal}). We have seen that the formation
of zonal jets is a basic phenomenon that can be related to the set of
invariants of the rotating shallow water dynamics.

For future work, it would be crucial to see whether the theoretical
predictions agree with experimental observations quantitatively, and
whether the effects of the extra invariant can be clearly resolved from
other mechanisms in the plethora of zonal jets phenomena. In
particular, we believe it important to develop our results for the
dynamics of magnetized plasmas; it would be very interesting to examine
the effects of the extra invariant on the formation of internal
transport barriers in fusion plasmas.

\begin{acknowledgments}

A.B.\ and P.W.\ wish to thank the Kavli Institute for Theoretical
Physics, University of California Santa Barbara, for hospitality and
support via the National Science Foundation Grant PHY05-51164; we are
grateful for beneficial discussions with the participants in two
Programs \emph{Physics of Climate Change} and \emph{Dynamo Theory} held
concurrently at the Institute. This paper is based upon work partially
supported by the National Science Foundation under Grant
DMS-0405905.

\end{acknowledgments}

\bibliography{My}
\end{document}